\documentclass[aps,twocolumn,superscriptaddress,showpacs]{revtex4}
\usepackage{graphicx}
\begin{document}
\newcommand{\tdna}{t^{\dagger}_{n\alpha}}
\newcommand{\tna}{t_{n\alpha}}
\newcommand{\tdnoa}{t^{\dagger}_{n+1,\alpha}}
\newcommand{\tnoa}{t_{n+1,\alpha}}
\newcommand{\tdng}{t^{\dagger}_{n\gamma}}
\newcommand{\tng}{t_{n\gamma}}
\newcommand{\tdnod}{t^{\dagger}_{n+1,\delta}}
\newcommand{\tnod}{t_{n+1,\delta}}
\newcommand{\tdag}{t^{\dagger}}
\newcommand{\al}{\alpha}
\newcommand{\ca}{\gamma}
\newcommand{\de}{\delta}
\newcommand{\taud}{\tau^{\dagger}}
\newcommand{\bfdelta}{ \mbox{\boldmath $\delta$} }
\newcommand{\sbfdelta}{\mbox{{\scriptsize\boldmath $\delta$}}}
\newcommand{\be}{\begin{equation}}
\newcommand{\ee}{\end{equation}}
\newcommand{\bea}{\begin{eqnarray}}
\newcommand{\eea}{\end{eqnarray}}
\newcommand{\half}{ \frac{1}{2} }   %\textonehalf\       \usepackage{textcomp}

\draft
\title{The Fate of the Two-Magnon Bound State in the Heisenberg-Ising Antiferromagnet}
\author{C. J. Hamer}
\affiliation{School of Physics, University of New South Wales,
Sydney NSW 2052, Australia}

\date{\today}
\begin{abstract}
The energy spectrum of the two-magnon bound states in the
Heisenberg-Ising antiferromagnet on the square lattice are calculated
using series expansion methods. The results confirm an earlier spin-wave
prediction of Oguchi and Ishikawa, that the bound states vanish into the
continuum before the isotropic Heisenberg limit is reached.
\end{abstract}
\pacs{05.30.-d,75.10.-b,75.10.Jm,75.30.Ds }
%(Submitted to  Phys. Rev. Lett.) }
\maketitle
%\newpage

\section{Introduction}
\label{sec1}

We consider the anisotropic antiferromagnetic Heisenberg model on a bipartite lattice
\begin{eqnarray}
H & = & J \sum_{<ij>}[S^z_iS^z_j + x(S^x_iS^x_j+S^y_iS^y_j)] \nonumber \\
 & = & J \sum_{<ij>}[S^z_iS^z_j + \frac{x}{2}(S^+_iS^-_j+S^-_iS^+_j)] \nonumber \\
 & \equiv & H_0 +xV
\label{eq1}
\end{eqnarray}
with S=1/2 spins interacting with their nearest neighbours.  

In the Ising limit $x = 0$, the system is N{\" e}el ordered, with spins
up ($S^z = +1/2$) on the A sublattice, let us say, and down ($S^z =
-1/2)$) on the B sublattice. The 1-particle `magnon' excitations
correspond to a single flipped spin on either the A sublattice, with
total spin $S^z = -1$, or on the B sublattice ($S^z = +1$). They have an
excitation energy $zJ/2$, where $z$ is the lattice coordination number.

Separated two-particle excitations then have energy $zJ$ in this limit;
but a two-particle excitation on neighbouring A and B sites has energy
only $(z-1)J$, with total spin $S^z=0$, forming a 2-particle bound
state.

When one takes the isotropic limit $x \rightarrow 1$, as is well known,
the energy gap for the single magnon states vanishes. The question then
is, do the 2-particle bound states survive in this limit?

It is well-known that two-magnon bound states exist for the isotropic
Heisebnberg {\it ferromagnet} - see for example the textbook discussion
by Mattis \cite{mattis1965}. In recent years, there have also been
considerable discussions of bound states in antiferromagnetic systems
with frustration or anisotropy \cite{extrarefs,trebst2000}. The question
still remains, however, whether there might also be bound states in the
simple, isotropic antiferromagnet. According to Mattis, this remained an
"unanswered question" in 1965 (Ref. \cite{mattis1965}, p. 166).

The question was investigated using spin-wave theory by Oguchi and
Ishikawa \cite{oguchi1973} in 1973. They used linear spin-wave theory
with some fourth-order interaction terms included. They found that the
two-magnon bound states merge into the continuum as one goes from the
Ising limit to the isotropic limit, so that none survive. This
calculation neglects many higher-order effects which might be important,
however, so it cannot be taken as definitive. Some numerical
calculations involving multi-magnon states have also been carried out
for the isotropic antiferromagnet on the square lattice. These include a
studies of the spectral weights using series expansions \cite{singh1995}
and quantum Monte Carlo simulations \cite{sandvik2001}, and a continuous
unitary transformation (CUTS) study of the spectrum \cite{schmidt2006}. None of these
works, however, have addressed the question of the bound states.  

In this paper we explore the fate of the two-magnon bound states for two
particular cases, the one-dimensional chain and the two-dimensional
square lattice. In one dimension, the model is exactly solvable, and the
answer is already known: we simply review the results. In two
dimensions, we use linked cluster methods \cite{oitmaa2006} to obtain
series expansions in $x$ up to order O($x^8$) for the bound-state
energies, and extrapolate the results to $x = 1$ using standard methods.
In summary, our results agree quite well with the spin-wave predictions
\cite{oguchi1973}. The bound states disappear into the 2-particle
continuum shortly before the isotropic limit $x \rightarrow 1$ is
reached. 

\section{The One-Dimensional Case}
\label{sec2}

 \begin{figure}
 \begin{center}
  \includegraphics[scale=0.4]{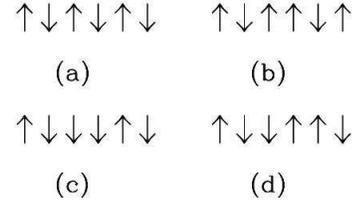}
   \caption{
Spin configurations on the linear chain in the Ising limit: a) N{\' e}el
ground state; b) a single spinon state; c) two adjacent spinons, or a
single magnon; c) two
spinons separated by one site, or two adjacent magnons.
}
 \label{fig1}
 \end{center}
 \end{figure}

The one-dimensional model, i.e. the linear chain, is a special case. The
model can be solved exactly using the Bethe ansatz \cite{yang1966}, and
exact expressions for the low-lying spectrum have been obtained
\cite{johnson1973}. The independent quasiparticles in this case are not $S^z =
\pm 1$ magnons, but $S^z = \pm 1/2$ `spinons', or domain walls. In the
Ising limit $x=0$, for instance, the ground state consists of
alternating spins $S^z = \pm 1/2$ on even or odd sites respectively, or
the reverse (Fig.\ref{fig1}a). A spinon or domain wall consists of a
neighbouring pair of identical spins in an otherwise alternating chain
(Fig.\ref{fig1}b). For periodic or anti-periodic boundary conditions,
spinons can only be created in pairs. Thus on an even lattice with
periodic boundary conditions, say, the lowest-lying excitations above the
ground state consist of a pair of spinons, which may be one lattice
spacing apart (Fig.\ref{fig1}c), two spacings apart (Fig.\ref{fig1}d),
or any number apart. Fig. \ref{fig1}c then corresponds to a single
spin-flip, and Fig. \ref{fig1}d to two neighbouring spin-flips, and so
on. The excitation energy in this limit, however, is $\Delta E = J$,
independent of the spacing.

Johnson, Krinsky and McCoy \cite{johnson1973} have obtained an exact expression for
the low-lying excitation energy at all $x$:
\begin{equation}
\Delta E = \frac{2K\sinh \nu}{\pi \cosh \nu} [(1-k^2\cos^2 q_1)^{1/2} +
(1-k^2 \cos^2 q_2)^{1/2}]
\label{eq2}
\end{equation}
where $\cosh \nu = 1/x$, $K$ is a complete elliptic integral of the
first kind whose nome is $q = E^{-\nu}$ and modulus $k^2$, and $ 0 \le
q_1,q_2 < \pi$ are two free parameters corresponding to the momenta of
the two domain walls in each sector. This spectrum is the same in the
$S^z = \pm 1$ and the (doubly degenerate) $S^z = 0$ sectors, so that the
two $S = 1/2$ spinons combine to form a degenerate $S=1$ triplet and $S
= 0$ singlet, at all couplings $x$ in the range. So in this case, the
state with two neighbouring flipped spins merely forms part of the
2-spinon continuum.

\section{The Square Lattice}
\label{sec3}

As the prime example of a bipartite system in two dimensions, we
consider the model on a square lattice. In this case no exact solution
is known, and so we have calculated numerical estimates for the energy
of the 2-magnon state using series methods \cite{oitmaa2006}. We perform an
Ising expansion, taking the Ising Hamiltonian $H_0$ in equation
(\ref{eq1}) as our unperturbed starting point, when the bound state
consists simply of a pair of flipped spins on neighbouring sites, as
discussed above. A perturbation series expansion in $x$ is then
calculated for the bound state energy, with $V$ in equation (\ref{eq1})
as the perturbation operator. 
As a technical point, we note that the bound
state lies in the same sector as the ground state, and hence a
`multiblock' diagonalization algorithm \cite{oitmaa2006} must be employed. 

\begin{figure}
 \begin{center}
  \includegraphics[scale=0.4]{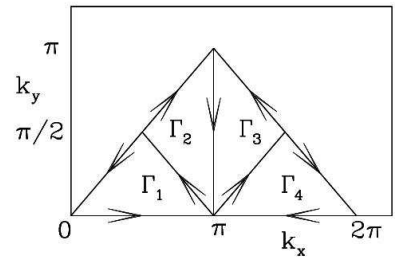}
   \caption{
Cuts through the Brillouin zone $\Gamma_1 - \Gamma_4$ corresponding to the four two-magnon
bound states.
}
 \label{fig2}
 \end{center}
 \end{figure}

Note also that there is one bound state configuration for each lattice
bond, making a total of four times as many configurations as for either
of the single-magnon states. Correspondingly, we obtain results for four
different paths $\Gamma_1, \cdots \Gamma_4$ in the Brillouin zone, as
shown in Fig. \ref{fig2}, whereas only the $\Gamma_1$ mode is independent for the
single-magnon state.

The calculations have been carried out through O($x^8$). Since
only even-order terms appear, this corresponds to only five series
coefficients at any fixed momentum. The leading order terms in the
dispersion relation for the bound state excitation energy are:
\begin{widetext}
\begin{eqnarray}
\epsilon({\bf k}) & = & 3-x^2\left[\frac{1}{4} +\frac{2}{3}\cos \frac{k_x}{2} \cos \frac{k_y}{2} +\frac{1}{12}(\cos k_x + \cos k_y) 
  +\frac{1}{6}(\cos \frac{k_x}{2} \cos \frac{3}{2}k_y + \cos \frac{3}{2}k_x \cos \frac{k_y}{2}) \right.
\nonumber \\
 & &  \left. +\frac{1}{6} \cos k_x \cos k_y  
   + \frac{1}{24}(\cos 2k_y +
\cos 2k_y)\right]
\label{eq3}
\end{eqnarray}
\end{widetext}
The complete series coefficients at selected momenta are listed in Table
\ref{tab1}.

\begin{figure*}[!tbp]
\includegraphics*[width=\columnwidth]{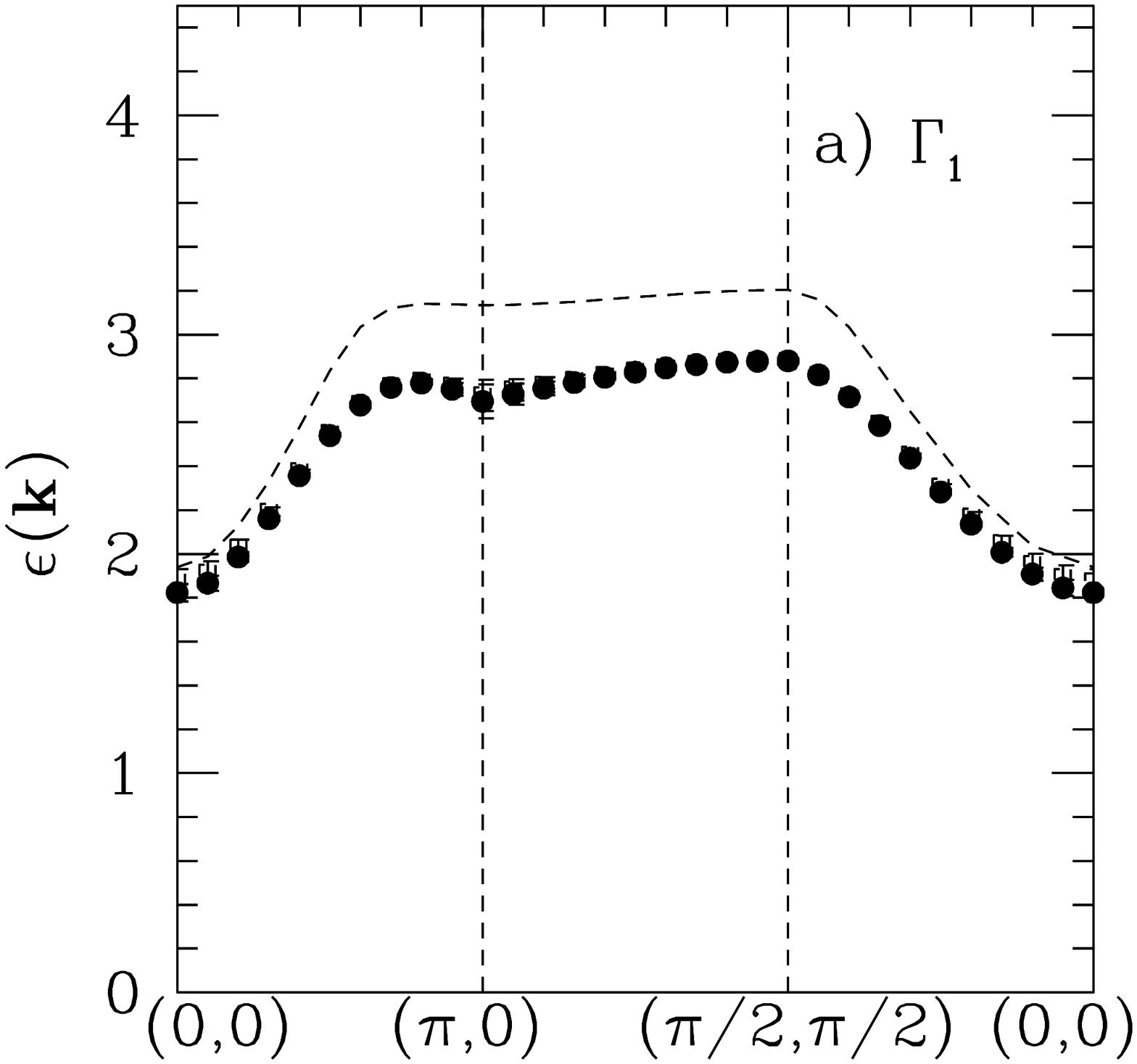}
\includegraphics*[width=\columnwidth]{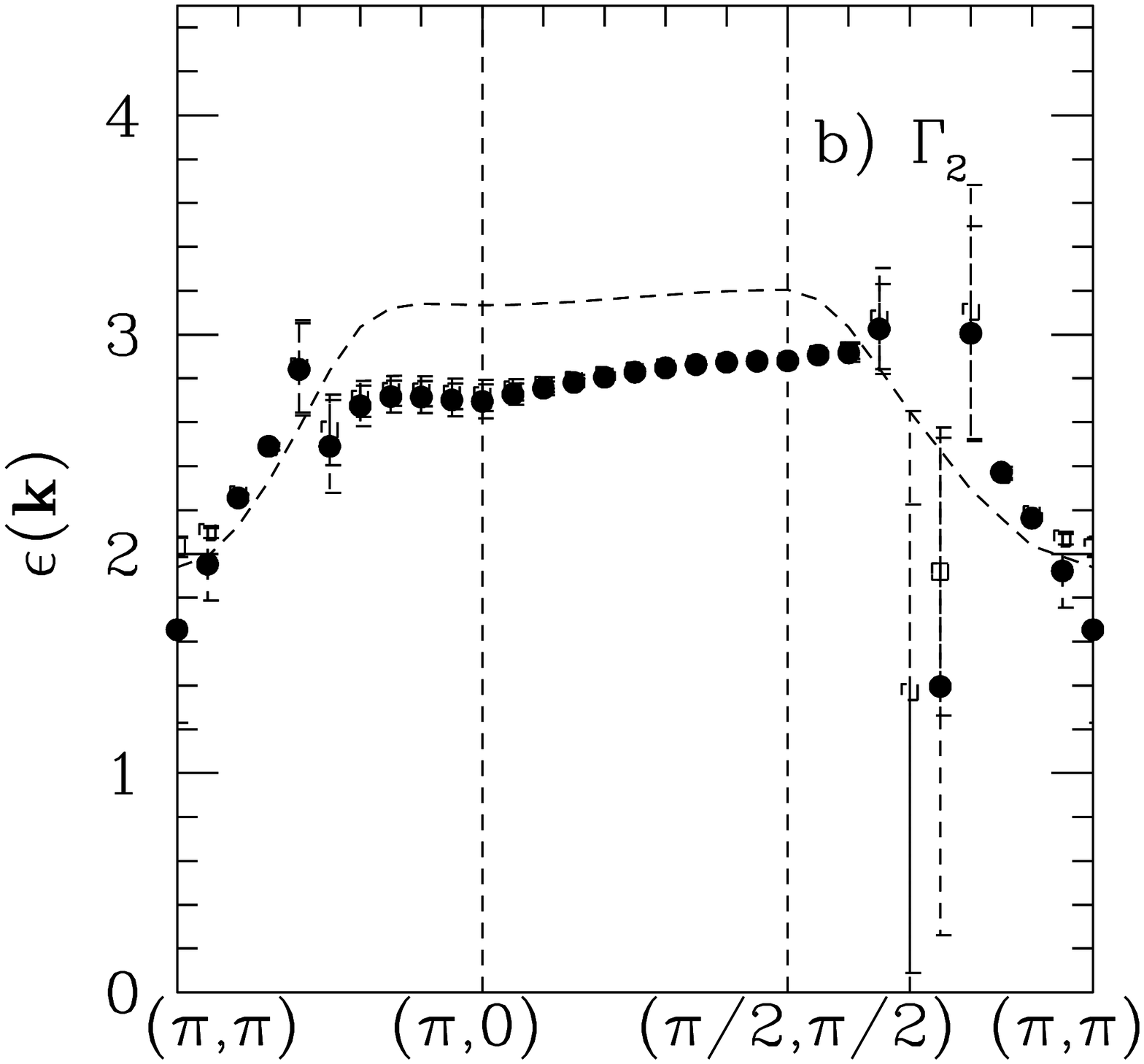}
\includegraphics*[width=\columnwidth]{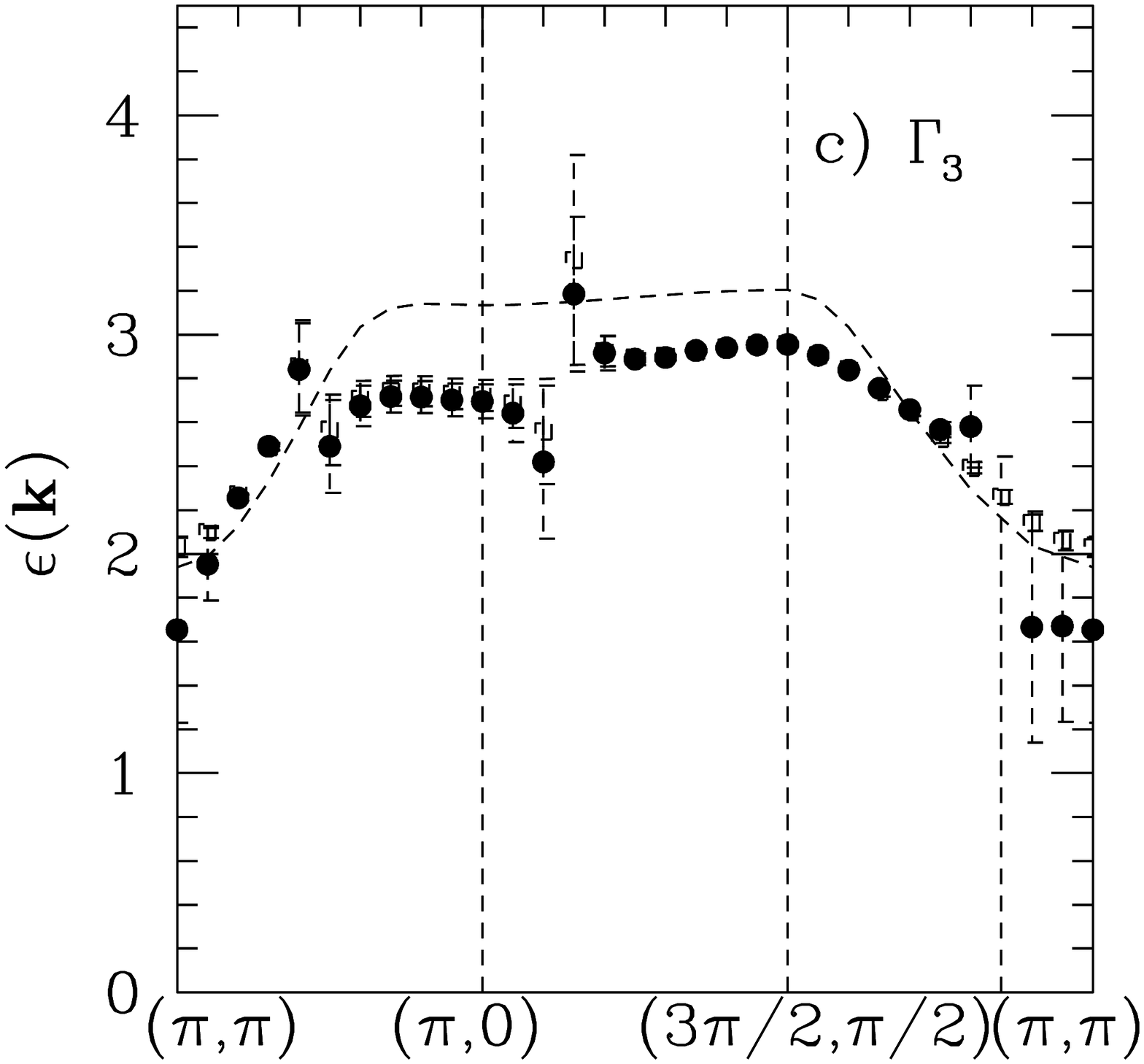}
\includegraphics*[width=\columnwidth]{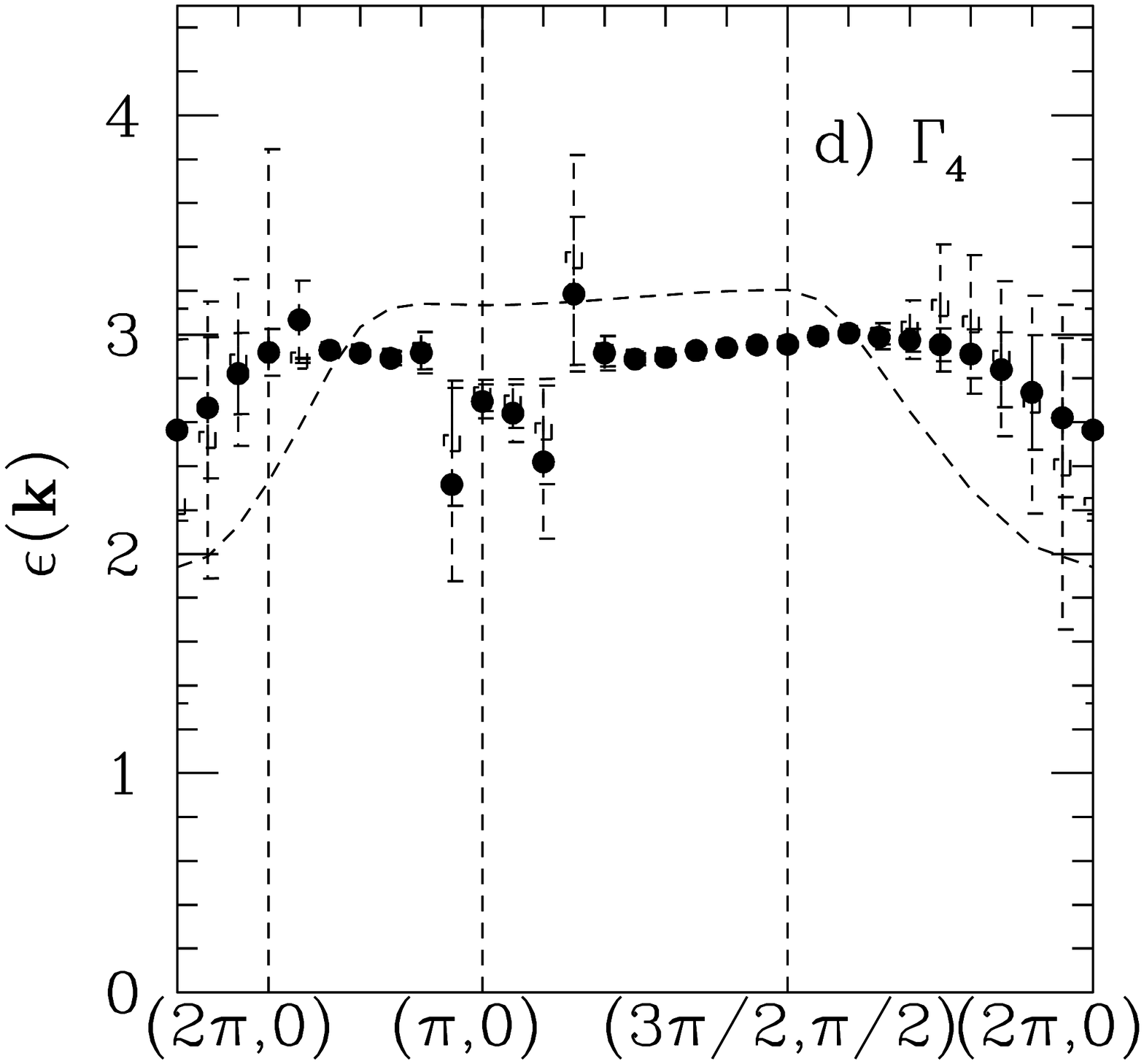}
   \caption{
Spectrum of the two-particle bound states $\Gamma_1 - \Gamma_4$ at $x
= 0.8$.
The dashed lines mark the lower edge of the 2-particle continuum.
}
 \label{fig3}
 \end{figure*}

Estimates of the bound-state energy as a function of $x$ were now
obtained using Pad{\' e} approximants to extrapolate the series. Since
the number of coefficients is small, the accuracy of the extrapolation
is also low. 
At smaller values of $x$, nevertheless, quite good estimates are
possible. For example, Figure \ref{fig3} shows dispersion relations for
the four bound-stste modes at $x = 0.8$, as compared with the lower
bound of the 2-particle continuum. It can be seen that all four modes 
remain bound below the intermediate plateau of the continuum, and only
near ${\bf k} = 0$ do three out of the four modes merge into the
continuum. The first mode appears to remain bound at all momenta. 

\begin{figure*}[!tbp]
\includegraphics*[width=\columnwidth]{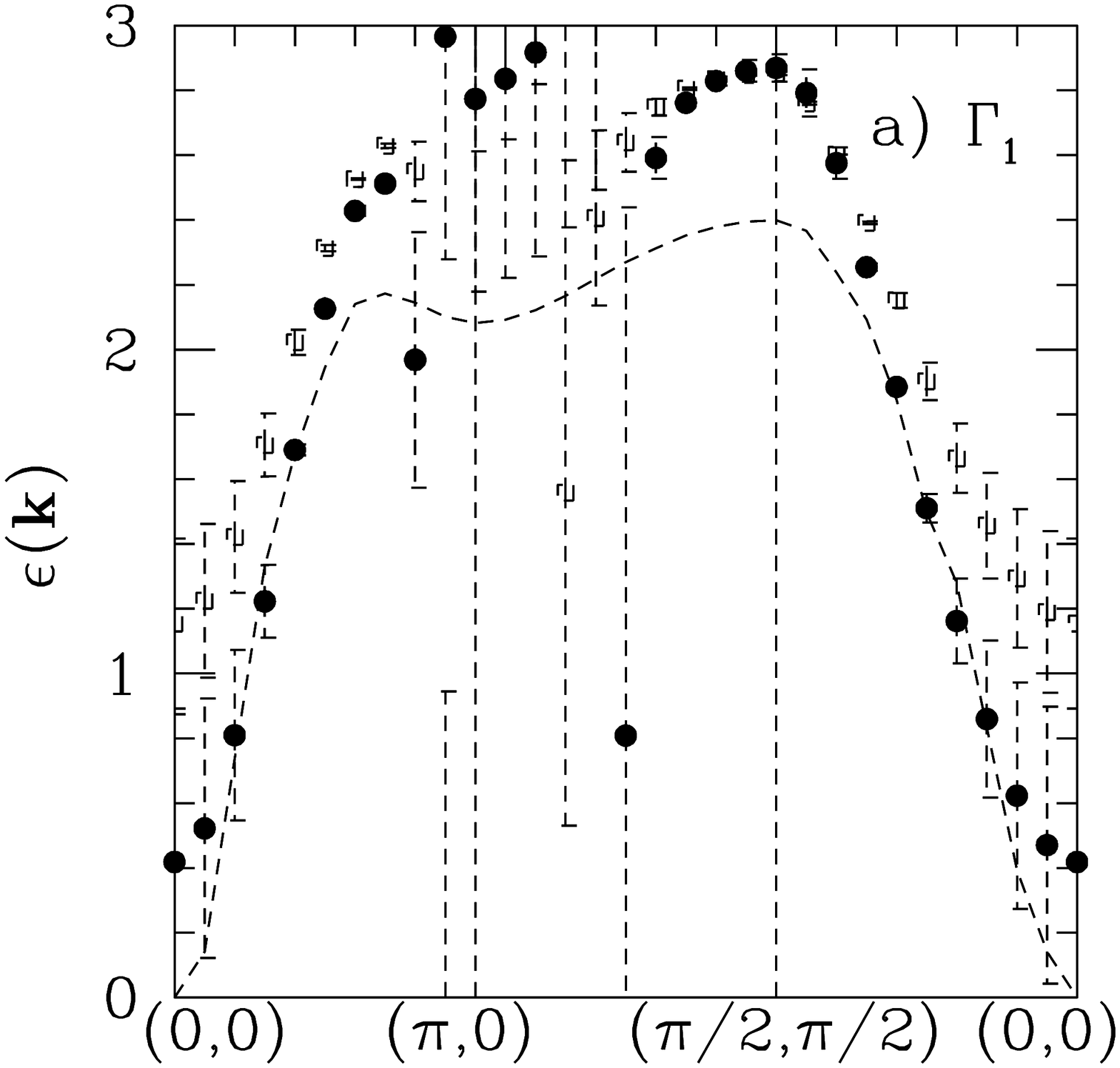}
\includegraphics*[width=\columnwidth]{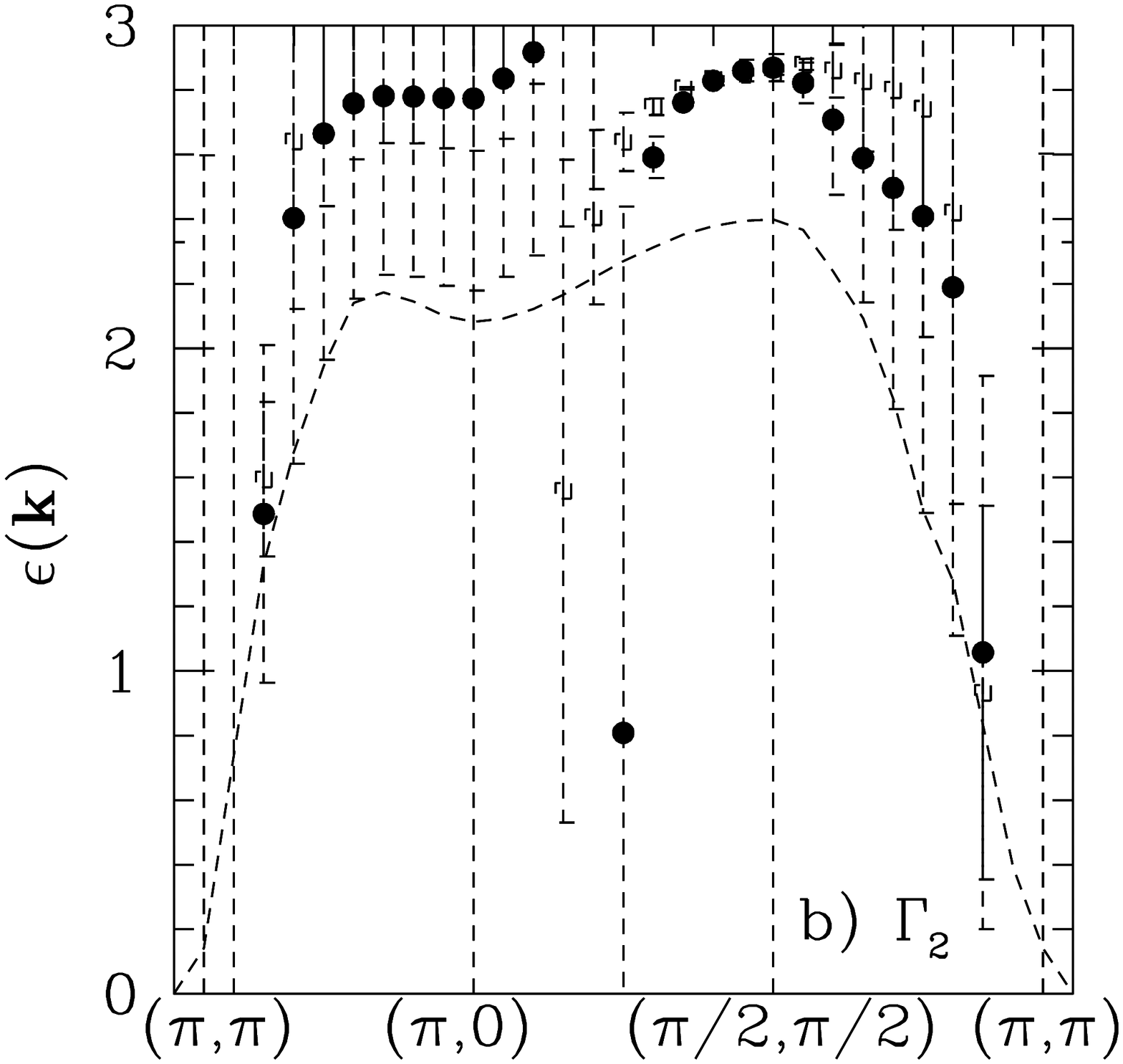}
\includegraphics*[width=\columnwidth]{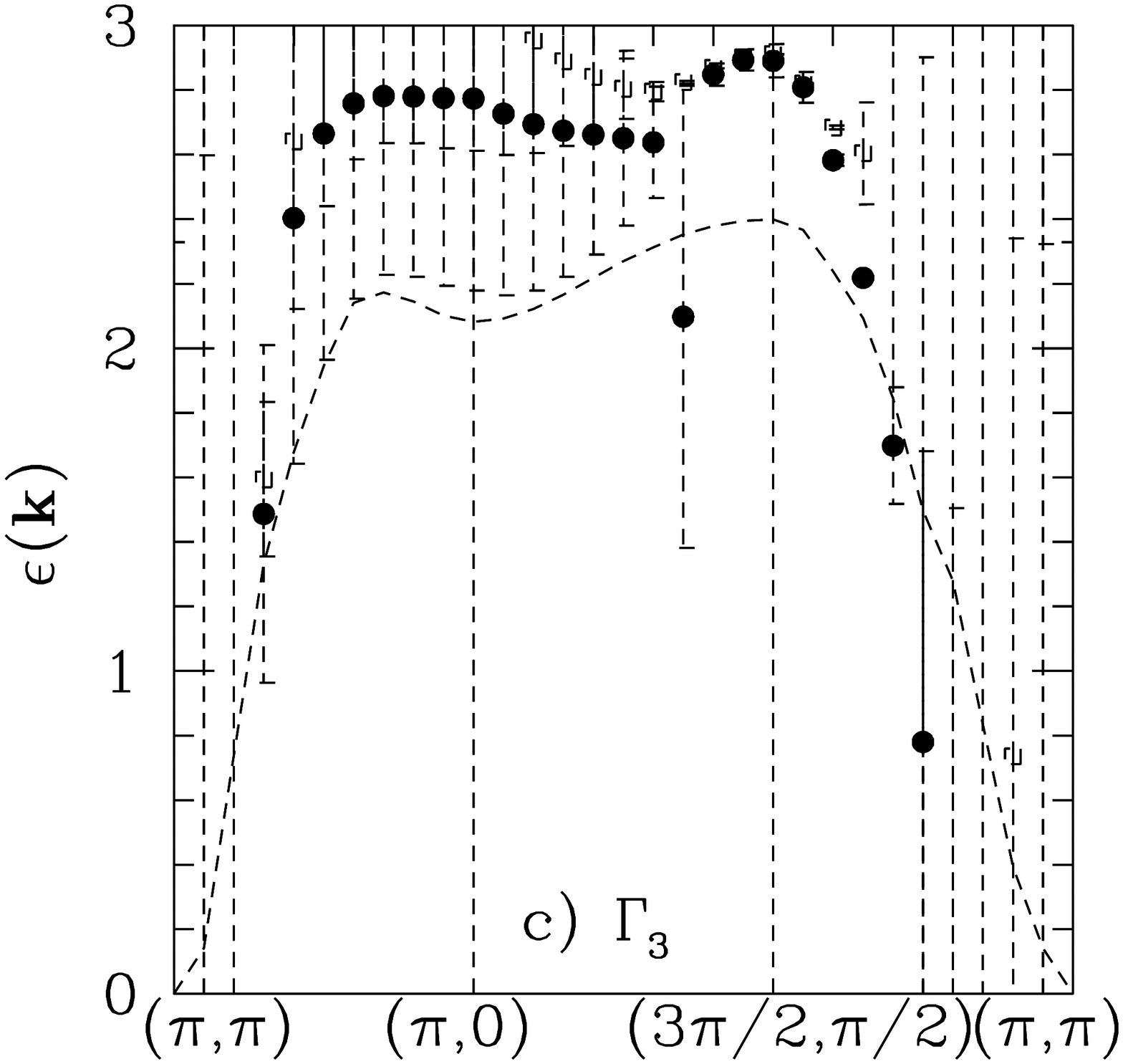}
\includegraphics*[width=\columnwidth]{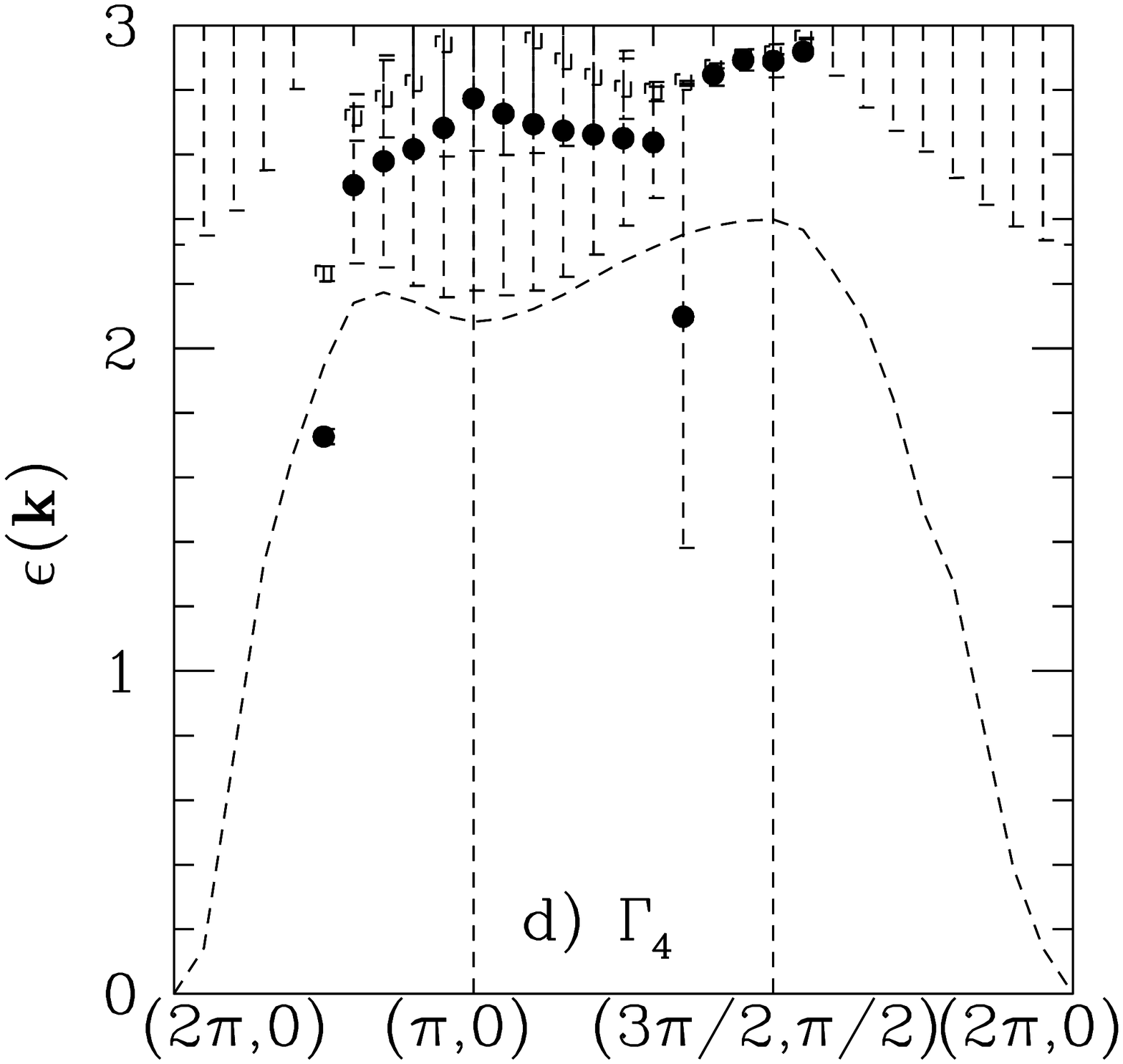}
   \caption{
Spectrum of the two-particle bound states $\Gamma_1 - \Gamma_4$ at $x =
1.0$. The dashed lines mark the lower edge of the 2-particle continuum.
}
 \label{fig4}
 \end{figure*}

At $x = 1.0$ it is a different story, as shown in Figure \ref{fig4}. In
short, it appears that none of the four modes remain bound at any
momentum. The nominal error bars are much larger in this case, firstly
because $x$ is larger, but also because we may expect some sort of
singular behaviour where the bound state merges with the continuum. The
clearest picture is obtained for the first mode, where near ${\bf k} =
0$ the estimates lie on the lower edge of the continuum, within errors.
There is an apparent levelling off at very small momenta, but this is a
common feature of Pad{\' e} extrapolations in Heisenberg-type models; if
a Huse transform \cite{huse} was performed before the extrapolation,
something much closer to the continuum behaviour would be expected. At
larger momenta, the estimates generally lie well above the continuum
lower bound.

These results may be compared with the spin-wave results of Oguchi and
Ishikawa \cite{oguchi1973}, who calculated the bound-state energies as
functions of $x$ for two specific momenta. At ${\bf k} = (0,0)$, they
found that the bound states merged with the continuum  in the range $0.7
\le x \le 0.95$; we find only one bound state remaining at $x=0.8$, in
agreement with their results. At ${\bf k} = (\pi/2,\pi/2)$, they found
the merger to occur at somewhat larger values $0.96 \le x \le 0.98$; we
find that all four states remain bound at $x = 0.8$, but have vanished
by $x = 1.0$, once more in agreement with their results. Of course, the
actual values for the energies have changed as higher-order terms are
added in, but not by a large amount.

\section{Conclusions}

We have used series expansion methods to calculate the energy of
two-magnon bound states in the anisotropic Heisenberg-Ising
antiferromagnet on the square lattice. We find that the bound states do
not survive in the isotropic limit, in agreement with the spin-wave
predictions of Oguchi and Ishikawa \cite{oguchi1973}. There are no bound
states in the linear chain model either. Hence one may extrapolate that
there will be no bound states in the isotropic Heisenberg
antiferromagnet on any bipartite lattice, in sharp distinction to the
ferromagnetic case. Oguchi and Ishikawa \cite{oguchi1973} have given
some qualitative arguments why this might be so.

\acknowledgments
I am grateful for useful discussions on this topic with Professors Rajiv
Singh, Goetz Uhrig, Oleg Sushkov and Jaan Oitmaa.
This work was supported by a grant from the Australian Research Council.
I am grateful for computational support from the Australian
Partnership for Advanced Computing (APAC) and the Australian Centre for
Advanced Computing and Communications (ac3).

\begin{widetext}
\begin{table}
\caption{
Ising expansion series coefficients in powers of $x$ for the excitation energy $\Delta E$ of the 2-particle bound state at selected momenta ${\bf k} =
(k_x,k_y)$.
}
\begin{ruledtabular}
\begin{tabular}{|c|c|c|c|c|c|}
 $(k_x,k_y)$     &     (0,0)         &   ($\pi$/2,0)           &      ($\pi$,0)         &     ($\pi/2,\pi/2$)  & ($\pi,\pi$)      \\
\hline
 0 &  3.00000000000000 &  3.00000000000000 &  3.00000000000000 &  3.00000000000000 &   3.00000000000000 \\
 2 &  -1.66666666666667 &  -0.804737854124365 &  -0.166666666666666 & -0.333333333333334  &  -0.333333333333333 \\
 4 &  0.299074074074084 &  0.412245678974668 &  -0.0712962962962898 &  0.403819444444453 &  -0.727546296296289 \\
 6 &  -2.21500154321004 &  -0.955337158046685 &  -0.371659167631400 &  -0.517003970550667 & -0.576196887860336  \\
 8 &  5.95191488216504 &  2.06716252423366 & 0.237274308271624 & 0.738026171850790 & -0.604532331023972    \\

\end{tabular}
\end{ruledtabular}
\label{tab1}
\end{table}
\end{widetext}

\end{document}